\documentclass{aa}
\usepackage{graphicx}
\begin{document}

\title{Properties of the \object{P Cyg} wind found using the Self Absorption Curve method}

   \author{
	G. Muratorio \inst{1}
\and
N. Markova \inst{2}
\and
	M. Friedjung \inst{3}
\and
	G.Israelian \inst{4}
         }

\offprints{M. Friedjung,\email{fried@iap.fr}}
\institute{
Laboratoire d'Astrophysique de Marseille, 2 Place le Verrier, FR-13248
Marseille Cedex 04, France\\
\email{muratorio@observatoire.cnrs-mrs.fr}
\and
Institute of Astronomy and Isaac Newton Institute of Chile Bulgarian Branch,\\
National Astronomical Observatory, P.O. Box 136, 4700 Smoljan, Bulgaria\\
\email{rozhen@mbox.digsys.bg}\\
\and
Institut d'Astrophysique, 98 bis Boulevard Arago, FR-75014
Paris, France\\
\email{fried@iap.fr}\\
\and
Instituto de Astrofisica de Canarias, Via Lactea, E-38200 La Laguna, Teneriffe, Spain\\
\email{gil@ll.iac.es}
   }

   \date{Received / Accepted}

   \abstract{
We have used the optical lines of N II and Fe III to study the wind of the
luminous blue variable \object{P Cyg}. This was performed by applying a version of 
 the Self Absorption Curve (SAC) method, involving few assumptions, to lines whose 
 flux can be measured.
A rather surprising result was obtained; 
the lines of more excited multiplets without blue shifted absorption  
components appear to be optically thick, while the lines of the most excited multiplets may show some indications of being optically thicker than  
the lines of less excited ones. Explanations of such effects are discussed,
including possible inhomogeneities in the wind.
   }

\authorrunning{Muratorio et al}
\titlerunning{Wind properties from Self Absorption Curve of \object{P Cygni}}

\maketitle

\section{Introduction}

The observational history of the hypergiant \object{P Cygni} begun about 400
years ago, when it was discovered by Willem Blaeu as a nova-like object. 
Because of its spectral and photometric characteristics the star is considered
to be the prototype of the P Cygni-type stars (PCT), as defined by Lamers 
(\cite{L86}), and as one of the prototypes of the giant eruption Luminous 
Blue Variables (Humphreys \cite{H99}). In addition, recent studies
(Markova et al. \cite{M01a}, de Groot et al. \cite{GSG}) showed that over the
last 15 years PCygni has behaved as a typical weak S Doradus variable in a
short S Dor phase (van Genderen \cite{vanGen01}). More information about this
enigmatic object can be found elsewhere (Israelian and de Groot \cite{IG99},
ASP Conf Ser. 233).

Stahl et al. (\cite{S93}) reported the presence of a number of pure emission
lines in addition to the dominant P Cygni-type lines in the optical spectrum
of \object{P Cygni}. Later Markova \& de Groot (\cite{MG97}) showed, comparing 
line-identification lists from various observational epochs,
that the optical emission spectrum of \object{P Cygni} was much richer and
intense in the mid nineties of the nineteenth century, than 60 years ago. In
particular, the authors noted that more than 70\% of the pure emission lines
appear to be, of recent origin. Among these are forbidden lines of Fe II, Ni
II, Ti II Fe III and N II; high excitation lines of Fe III and  N II and lines
of low and medium excitation of Si II.  If not due to observational selection
caused by the continuous improvement of the signal-to-noise ratio of
photographic spectra during the last century, this result might indicate the
presence of a very long-term variation in the wind+photospheric properties
of the star. In this context, it seems worthwhile to try to obtain  
additional information on the nature and the origin of P Cygni's emission 
spectrum  using methods not used up to now, such as for instance the SAC 
(Self-Absorption Curve) method developed by Friedjung and Muratorio 
(\cite{FM87}). This method involves a semi-empirical analysis of emission
line spectra of complex atoms and ions, without assuming detailed models
for the objects emitting these spectra. By means of the SAC method valuable
information (e.g. the self absorption effects, the level population laws  
and population anomalies) has been obtained for the Fe II emission spectrum
of many stars of different types, such as  AG Car (Muratorio and Friedjung
\cite{MF88}), the VV Cep star KQ Pup (Muratorio et al \cite{M92}), while
procedures for using the method is described by Baratta et al (\cite {B98}).
Use of the SAC method can require however a few hypotheses about the relative
populations of levels inside the same spectroscopic term and the distribution
of self-absorption relative to that of the emission of different lines in
different parts of the line emitting region. The second of these hypotheses
was in any case not needed in the present work.

The main purpose of our study is to apply the SAC method to the Fe III and
N II emission spectra of \object{P Cygni}. In this way we can hope to gain additional
information about the nature and the origin of these spectra and thus to get
a deeper insight into the physics of the star itself. The observational
material is described in Section 2, where a number of  problems 
concerning the derivation of the net emission equivalent widths of the  
studied lines are also discussed and resolved. Section 3 presents the results
obtained through the SAC analysis while Section 4 deals with the
interpretation of these results.

\section{Observations and data reduction. EW measurements}

The present analysis is based on one spectrum  taken at the Heidelberg
Observatory on August 9 1991 with a fibre echelle spectrograph attached to a
telescope of the 70 cm class (Stahl et al \cite{S95}). An EEV CCD with 
1252 $X$ 770 pixels of 22 $\mu$ size was used as a detector. The spectrum covers
an interval of 2\,700\AA, from $\lambda$4001 to $\lambda$6773, with a 
spectral resolution of 12\,000 and a signal to noise ratio of 333  (Stahl
et al. \cite{S93}). In cases where measurements were doubtful and also  
in order to confirm the presence of faint lines, spectra obtained with the
AURELIE spectrograph of the Haute Provence Observatory in July 1997, were
examined. No variatiations were found, at least for the lines studied. The
spectral resolution of these AURELIE spectra is 10,000 and the signal to noise
ratio similar to that of the spectrum used in the present work.

The main requirement of  the SAC method to be applied is the presence  a  
large number of emission lines of different multiplets of the same ion which
form in the same region of the stellar envelope. \object{P Cygni} shows an
optical spectrum that is quite rich in emission lines (permitted and
forbidden), including in particular many lines of N II and Fe III. The
permitted transitions from less excited levels are accompanied by blueshifted
absorption components, while the lines from more excited levels are,
like the forbidden lines, purely in emission  (Stahl et al. \cite{S91}, Stahl
et al. \cite{S93}, Markova and de Groot (\cite{MG97}). To increase the number
of lines in our sample we decided to analyse in addition to the permitted
lines with a pure emission profile also those with a P Cygni-type
profile. Thus the sample of lines, we were able to analyse by the SAC method,
consists of 30 lines from  7 multiplets of N II and 42 lines from 11  
multiplets of Fe III. The multiplets as well as the distribution of the lines
by multiplets are given in Columns 1 and 6 of Table 1.

To construct the SAC of a given multiplet one needs to know  the lower-level
statistical weight, $g$, the oscillator strength, $f$ and the deredenned
emission flux, F$_{\lambda}$, 
of each line from this multiplet. In the  present analysis we have used net
emission equivalent widths (i.e.  equivalent widths corrected for photospheric
absorption), which we converted into line fluxes using a deredenned continuum
calculated from  B = 5.14 and V = 4.72 and deredenned with
E(B-V)$_{\rm P Cyg}$ = 0.63. In deriving the net emission equivalent widths
(W$_\lambda$) of the  lines from their observed profiles, we 
were confronted by the following problems: \\
(i)  possible contamination  by photospheric absorption;\\  
(ii) overlapping between wind absorption and emission in the case of lines
with a P Cygni-type profile; \\
(iii) blending effects between pure emission lines which are not well-resolved
in the spectrum. \\

The correction for photospheric absorption was performed by subtracting a
synthethic spectrum, derived, using model calculations of Israelian
(\cite{I95}), from the observed spectrum. The spectrum, obtained in this way,
was then measured for the equivalent widths of the lines. It must be noted
here, that we were not fully convinced that such corrections had to be
done, nor whether they were valid. This is because  Markova et al.
(\cite{M01b}) showed that the wind of \object{P Cygni} is probably sufficiently
optically thick for the star to form a permanent pseudo-photosphere. If this
is the case, the observed profiles do not need to be corrected for
photospheric absorption. However, Markova and de Groot (\cite{MG97}) found
that the emission peaks of the \object{P Cygni} profiles of Balmer, He~I and N~II lines
in their 1990 spectra, were red-shifted with respect to the systemic velocity
of the star, V$_{\rm sys}$ { = -22 km s$^{-1}$}, a result that can be easily
explained if one believes that these peaks are distorted by photospheric
absorption. In addition, Israelian (\cite{I95}) argued that the complex
structure of [Fe~II] forbidden lines in P Cyg's spectrum is due to blending
effects caused by weak photospheric lines. {The results outlined above can
be reconciled, if one suggests that the wind opacity can change significantly,
a conjecture that seems quite reasonable. In fact Markova et al. (\cite{M01a})
showed that the mass loss rate of \object{P Cygni} can vary within 26\% 
on a timescale of at least 600 days. In any case the photospheric
correction has a small effect on our results}  

The contribution of wind absorption to the P Cygni-type profiles was
accounted for by means of two methods, both of which turned out to give
similar results as far as SAC slopes are concerned. The first way we
proceeded was to measure  the flux of the red part of each profile,
integrating the emission flux redward of the emission peak, and multiplying
the value obtained by 2. This aproach seems to be appropriate for our
case since emission components are expected to be symmetric with respect to
their peak intensity (Markova \& de Groot \cite{MG97}). The second way we  
proceeded was to fit various Gaussians to the observed \object{P Cygni} profiles.
It turned out that three Gaussions of different FWHM  (Full Width at Half Maximum)

\begin{figure*}
\centering
\includegraphics[angle=-90,width=8.5cm,bb=55 100 530 780,clip]{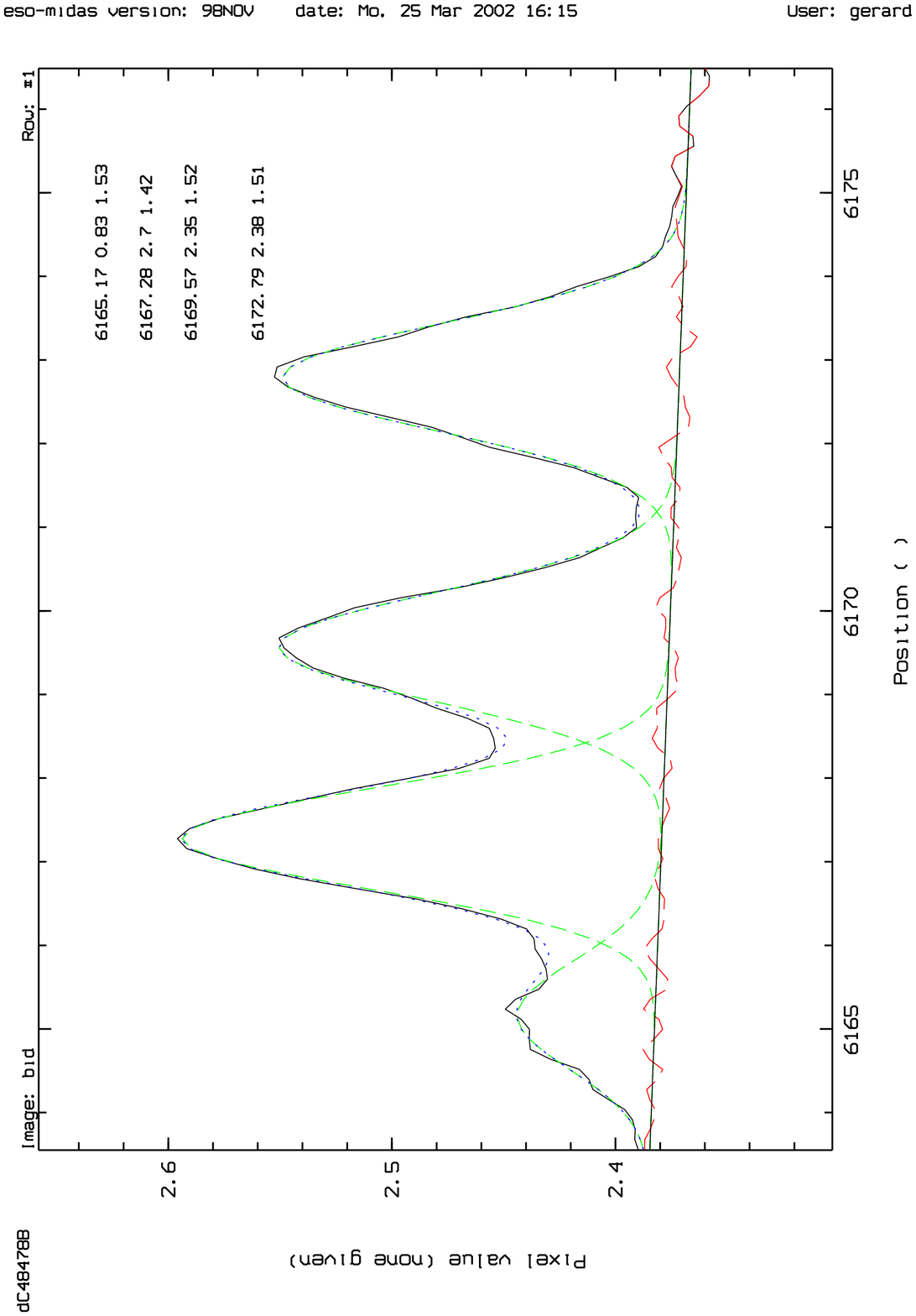}
\includegraphics[angle=-90,width=8.5cm,bb=55 100 530 780,clip]{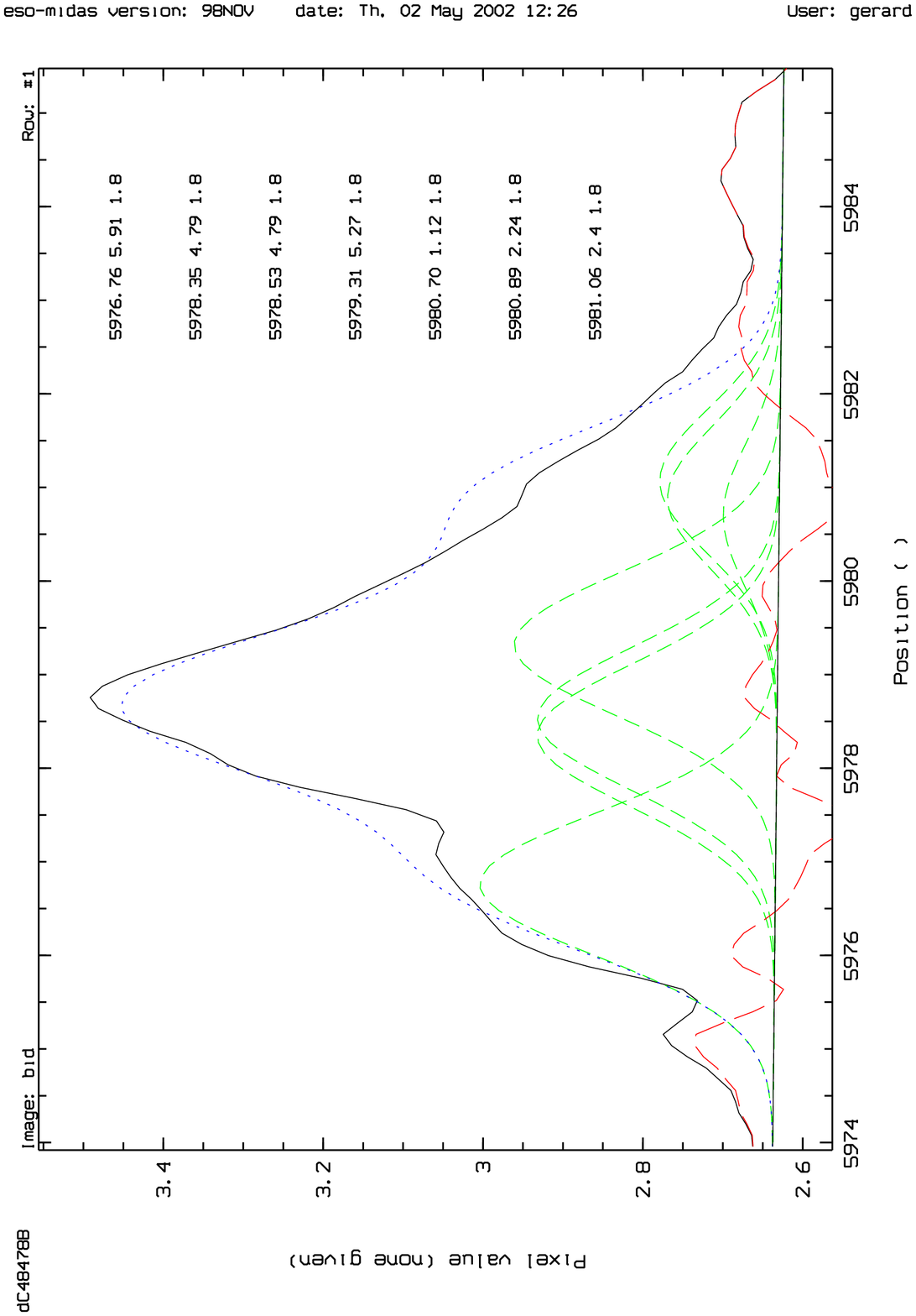}
\caption{Two examples of line fitting using gaussians : left side, fitting of
four weakly blended lines. right side, fitting of seven severely blended lines,
six of which are from multiplet `` 756'' whose SAC is displayed on fig.2.}
\label{Fig1}
\end{figure*}

-- two centered at V$_{\rm sys}$ for the emission and one for the absorption --
are usually enough, as a first approximation, to obtain a satisfactory fit.
The contribution of absorption at velocities of more than V$_{\rm sys}$ in this
type of fit is small.\\

To separate and measure blends between pure emission lines we used 
Gaussian fits to the profiles. This fitting was performed taking into
account the velocity and FWHM of the lines. Blending was also considered taking
into account the components's location on the self-absorption curve. Two examples
of blend separations are shown in Fig. 1.
The numbers given in the upper
right hand corner of the figure denote the wavelength of the central position,
the flux and the Gaussian width (in \AA) of each component. Let it be noted
that not only the differences in line flux, represented by a dashed line,
between the fit (dotted line) and the observations (solid line), are very
small,  but also that the central velocities and the FWHM of the three N II
lines (multiplet  36), as determined by the fit, are in perfect agreement with
each other as would be expected if the lines were formed in the same region of
the wind. The equivalent widths of the single pure emission lines were
measured, fitting their profiles with a Gaussian.

\section{Application of the SAC method to P Cyg}

At the beginning of this section we would like to note that up to now the 
SAC method has, in all published papers except for conference proceedings,
been applied to study the emission spectrum of the Fe II ion only. In  
the present study, we shall for the first time try to apply this method to
the spectrum of Fe III and N II.

In our present work we have firstly tested that the points of the graphs  
of $\log\frac{{F\lambda^{3}}}{{gf}}$ against $\log(gf\lambda)$ for each 
multiplet lie near a curve. The slopes of the curves were then compared for 
different multiplets. The graphs with linear least square solutions for
different multiplets are plotted in Figs. 2 and 3

Among these plots are two for different groups of very highly excited Fe III
lines, those with lower to upper excitation potentials from 22.54 to 24.59 eV
being called ``multiplet 705'' and those with the corresponding excitation
potentials from 23.61 to 25.68 eV being called ``multiplet 756''.
The results of the least squares solutions, where at least three lines are 
present, are given in Table 1.

The first thing we see is that the vertical dispersions of the points around
the linear solutions are in most cases not very large in the same multiplet.
The RMS (root mean square) deviation is less than 0.4 for all multiplets
except multiplets 756 of Fe III and 46 of N II. Secondly, the slopes of the
graphs, found by linear least squares solutions, are still large for more
excited multiplets, indicating that their lines are optically thick, even
though their blue shifted absorption is not detectable. This is still true if
only the more significant solutions are taken, for which the values of log
$(gf \lambda)$ of some lines in the same multiplet differ by more than 0.5.
There are even signs of a slope increase for the most excited multiplets. The
range of log $(gf \lambda)$ values hardly changes between the least and the
most excited multiplets of N II, while in the case of Fe III, the 
log $(gf \lambda)$ values are however considerably less for the first 3 
multiplets. The slopes and other relevant information are listed in Table 1.

\begin{table}
\caption[]{Results from the SAC analysis}
\begin{tabular}{llllll}
\hline
ion +    &excit&potent&   SAC & RMS & lines \\
multiplet&lower& upper& slope &&\\
\hline
FeIII 4	 & 8.2 & 11.1 & -0.40   & 0.2 & 4 \\
FeIII 5	 & 8.6 & 11.1 & -0.16   & 0.2 & 8 \\
FeIII 68 & 14.1& 16.5 &  0.00   & 0.2 & 3 \\
FeIII 113& 18.2& 20.6 & -0.35   & 0.2 & 7 \\
FeIII 114& 18.4& 20.6 &         &     & 2 \\
FeIII 115& 18.7& 20.9 &         &     & 2 \\
FeIII 117& 18.8& 20.9 & -0.42   & 0.3 & 3 \\
FeIII 118& 20.6& 23.6 & -0.44   & 0.2 & 4 \\
FeIII 119& 20.6& 23.7 &         &     & 2 \\
FeIII 705& 22.5& 24.6 &         &     & 2 \\
FeIII 756& 23.6& 25.7 & -0.80   & 0.4 & 5 \\
         &     &      &         &     &   \\
NII 3	 & 18.5& 20.7 & -0.44   & 0.1 & 6 \\
NII 5	 & 18.5& 21.2 & -0.62   & 0.2 & 6 \\
NII 19	 & 20.7& 23.1 & -0.70   & 0.2 & 3 \\
NII 20	 & 20.7& 23.2 & -0.21   & 0.1 & 4 \\
NII 28	 & 21.2& 23.2 & -0.81   & 0.15& 4 \\
NII 36	 & 23.1& 25.1 & -0.81   & 0.1 & 3 \\
NII 46	 & 23.2& 25.2 & -0.86   & 0.40& 4 \\
\hline
\end{tabular}
\end{table}

\begin{figure*}
\centering
\includegraphics[angle=-90,width=8.5cm,bb= 55 100 530 780,clip]{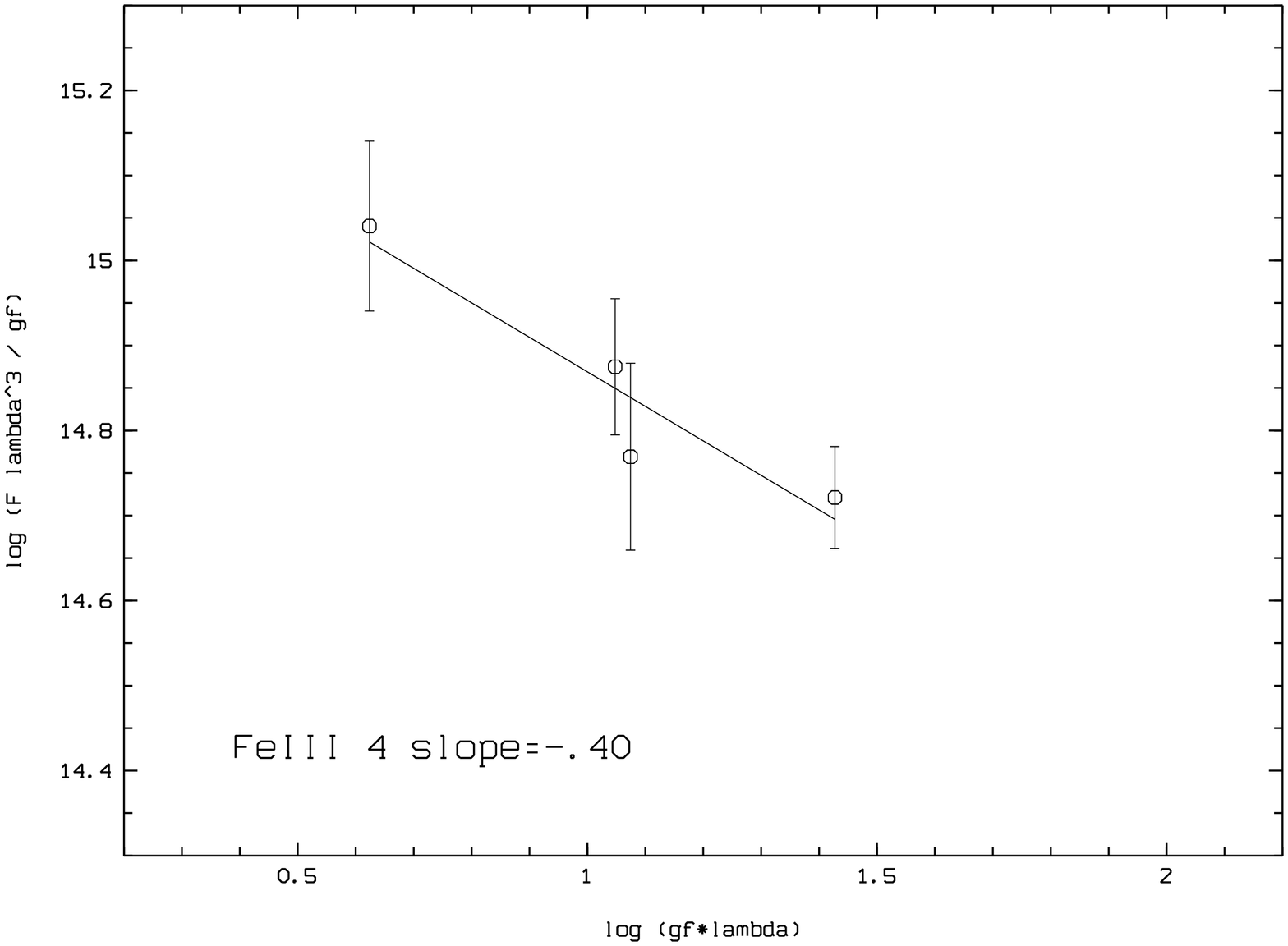}
\includegraphics[angle=-90,width=8.5cm,bb= 55 100 530 780,clip]{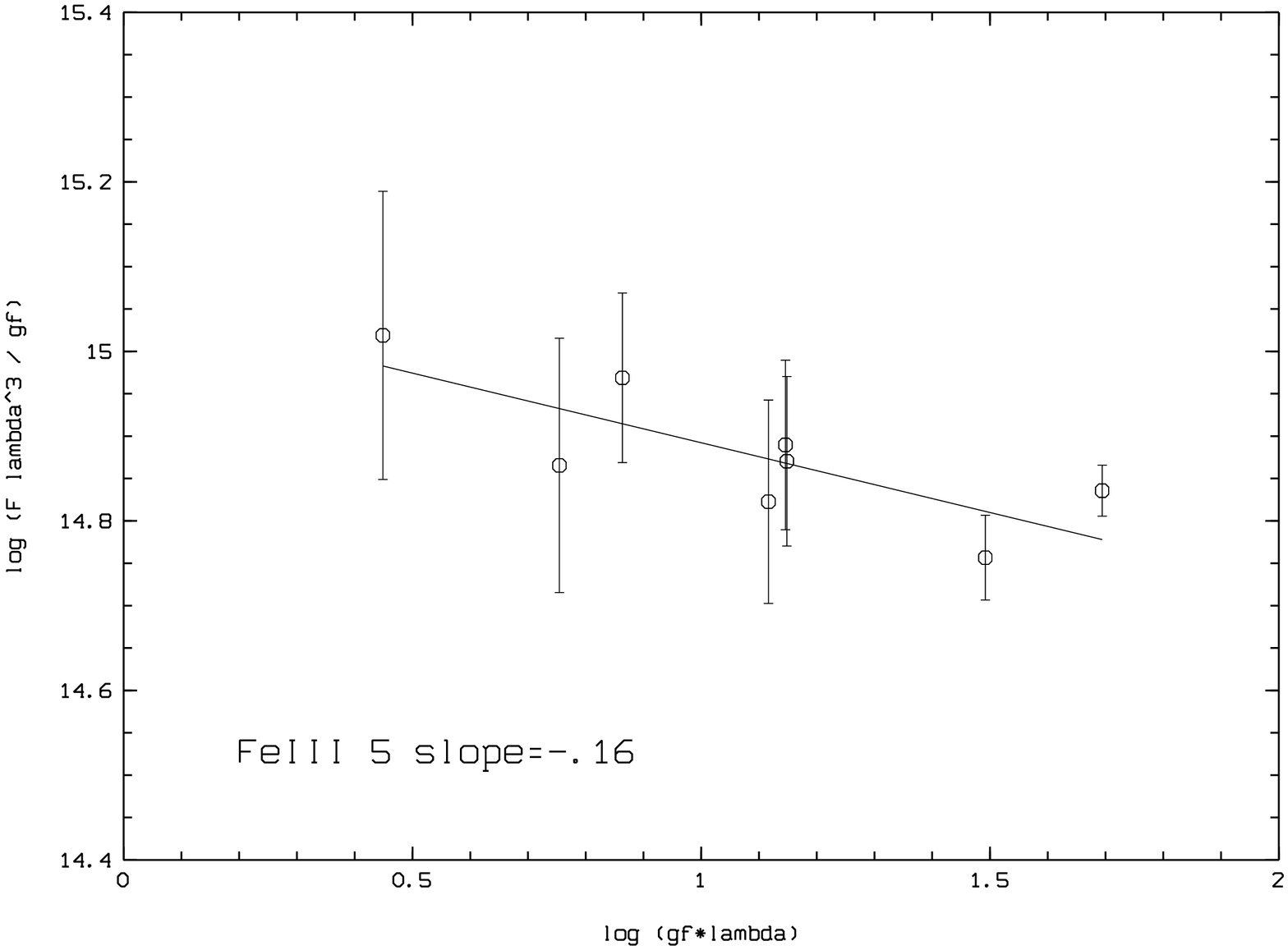}
\includegraphics[angle=-90,width=8.5cm,bb= 55 100 530 780,clip]{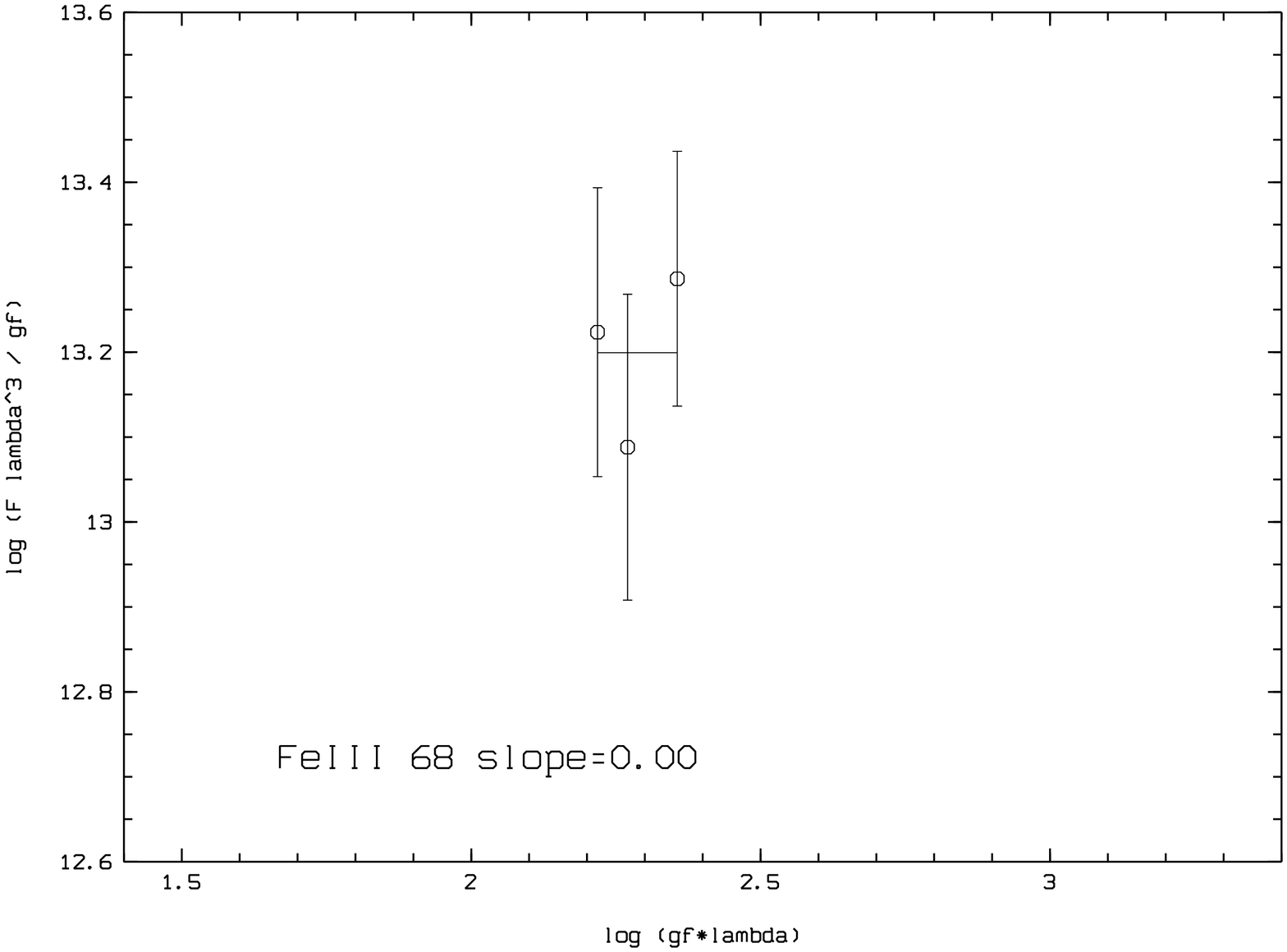}
\includegraphics[angle=-90,width=8.5cm,bb= 55 100 530 780,clip]{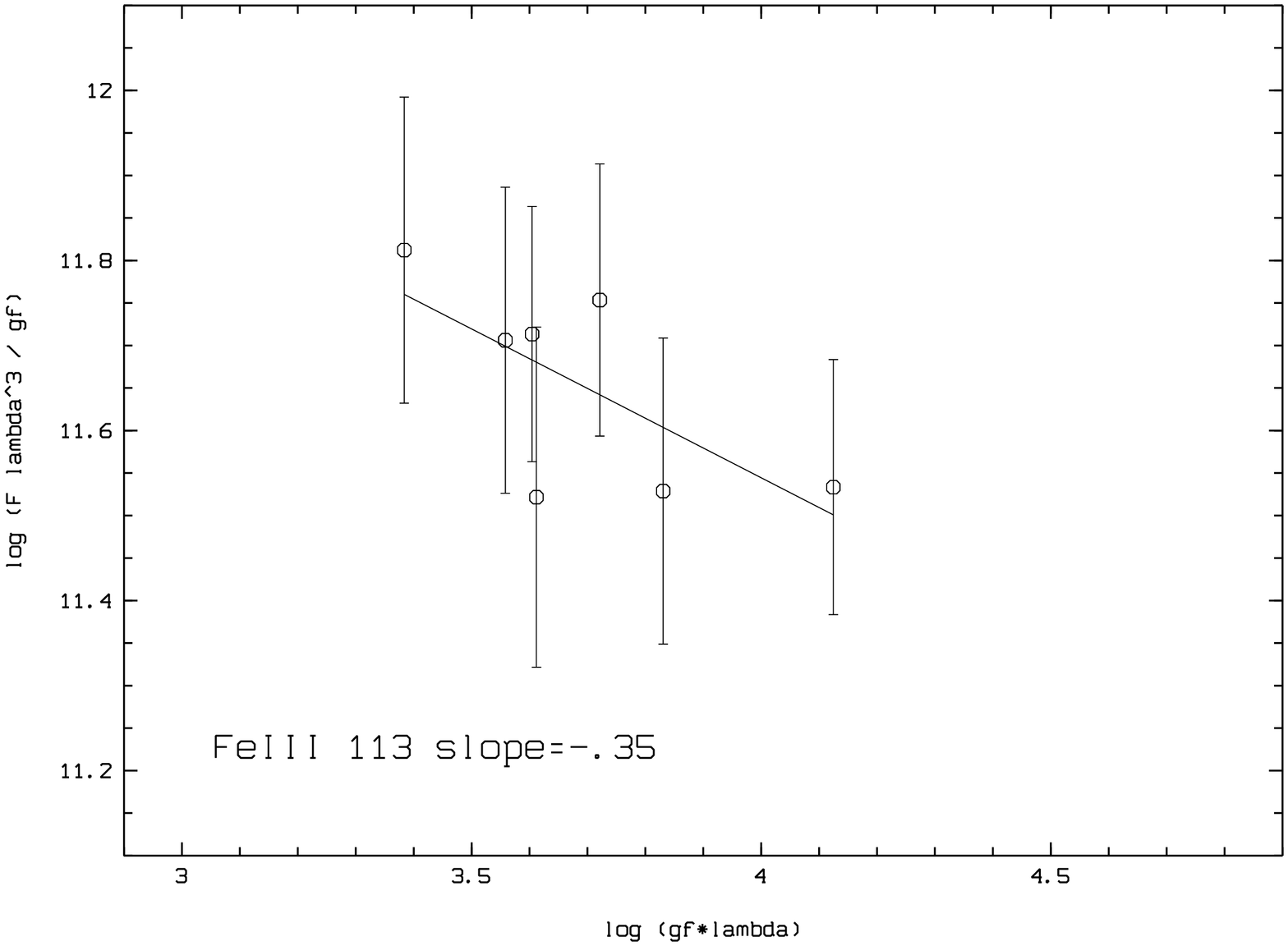}
\includegraphics[angle=-90,width=8.5cm,bb= 55 100 530 780,clip]{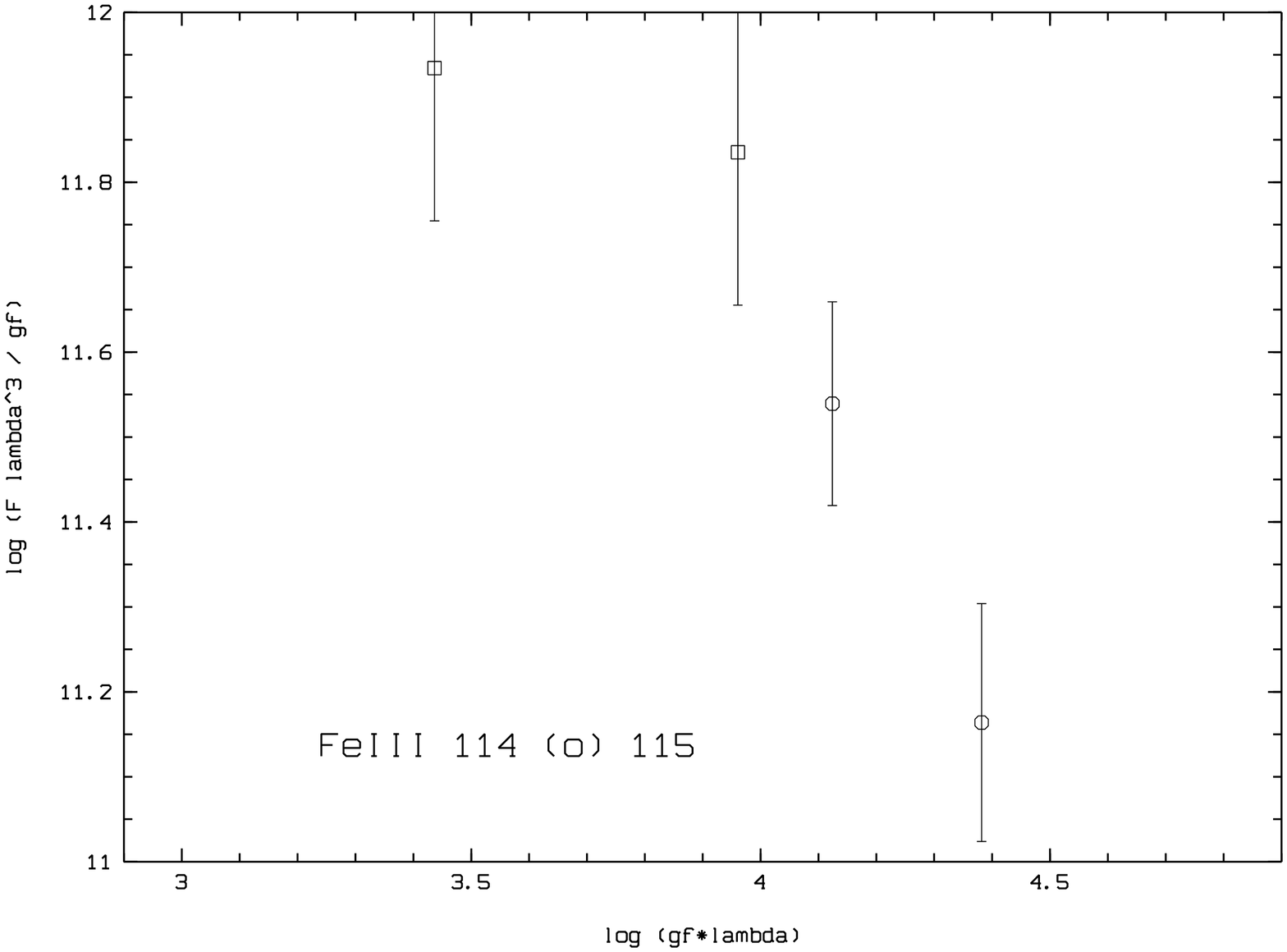}
\includegraphics[angle=-90,width=8.5cm,bb= 55 100 530 780,clip]{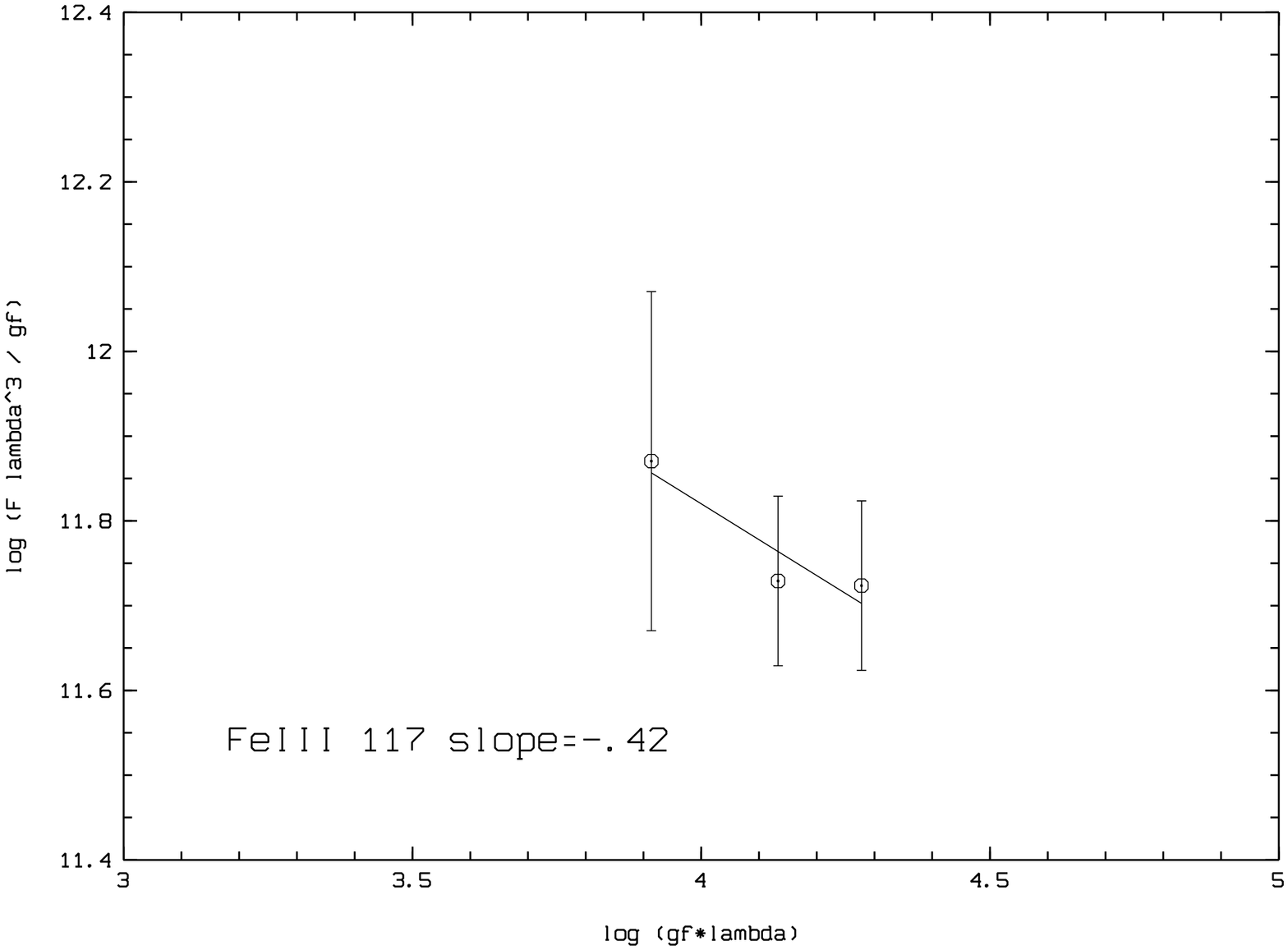}
\includegraphics[angle=-90,width=8.5cm,bb= 55 100 530 780,clip]{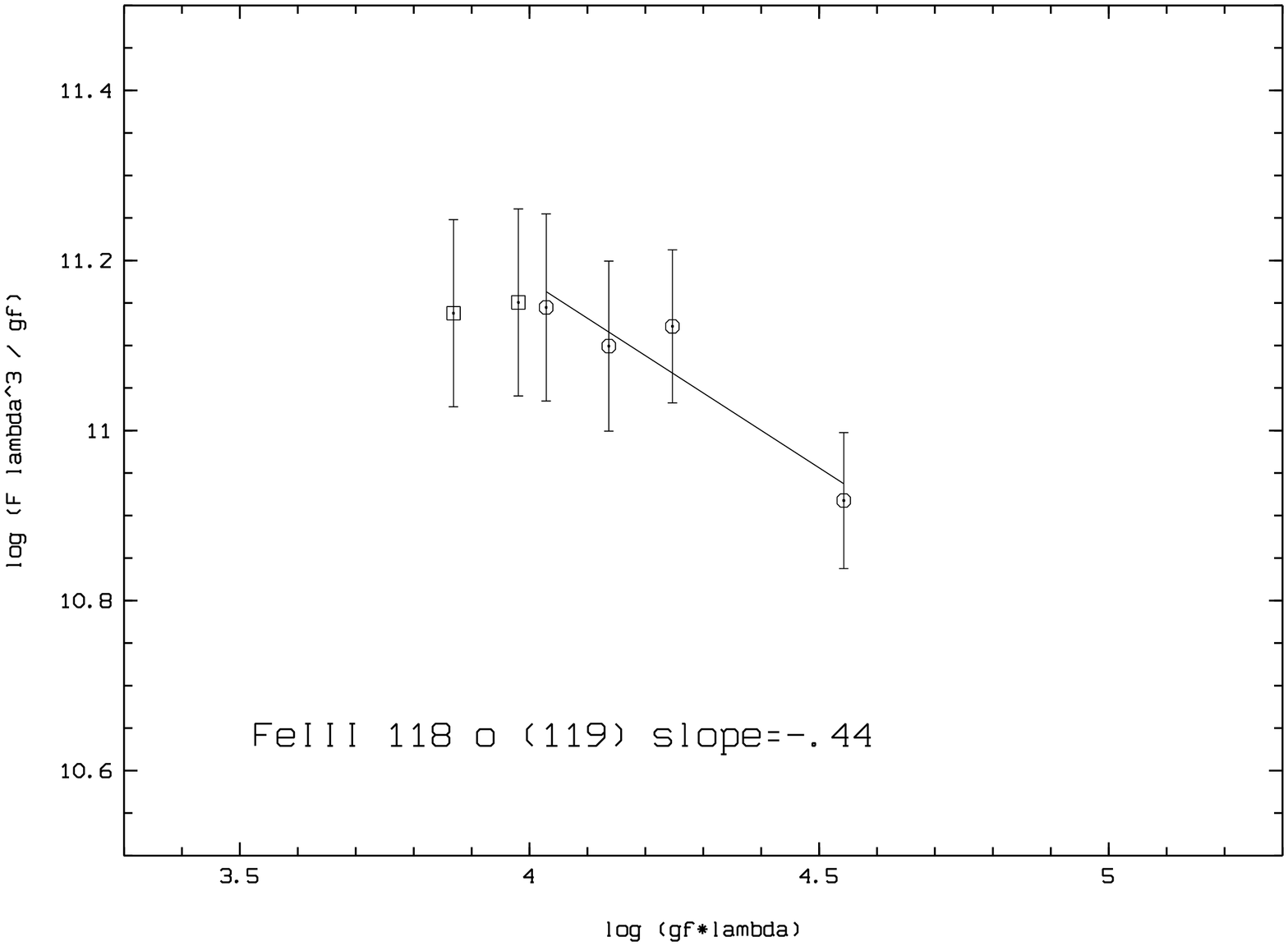}
\includegraphics[angle=-90,width=8.5cm,bb= 55 100 530 780,clip]{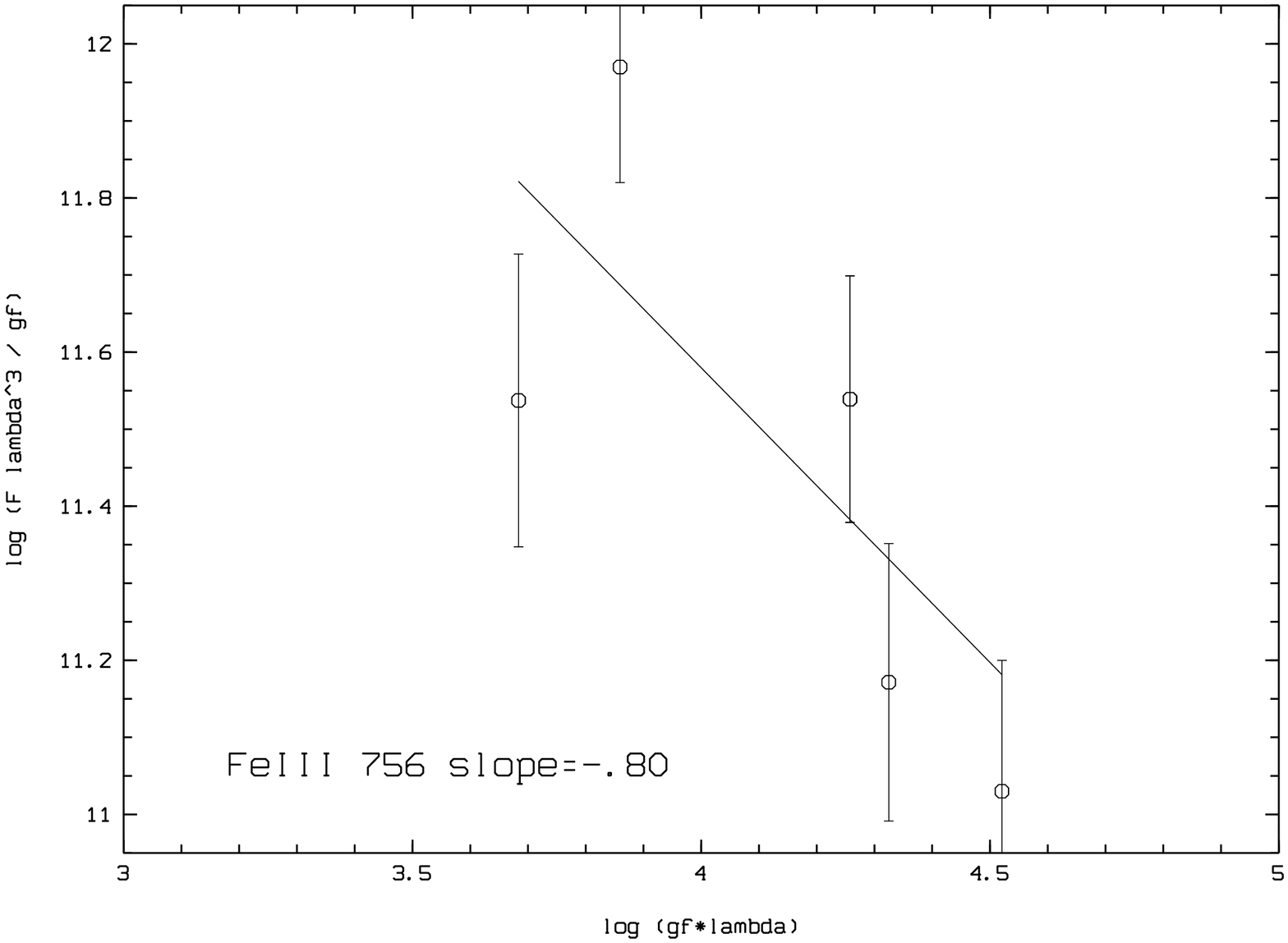}
\caption[width=\textwidth]{Plots of $\log\frac{{F\lambda^{3}}}{{gf}}$ against $\log(gf\lambda)$  (SAC) for FeIII multiplets 4,5,68,113,114,115,117,118,119 and "756"}
\label{Fig2}
\end{figure*}

\begin{figure}
\centering
\includegraphics[angle=-90,width=8.5cm,bb= 55 100 530 780,clip]{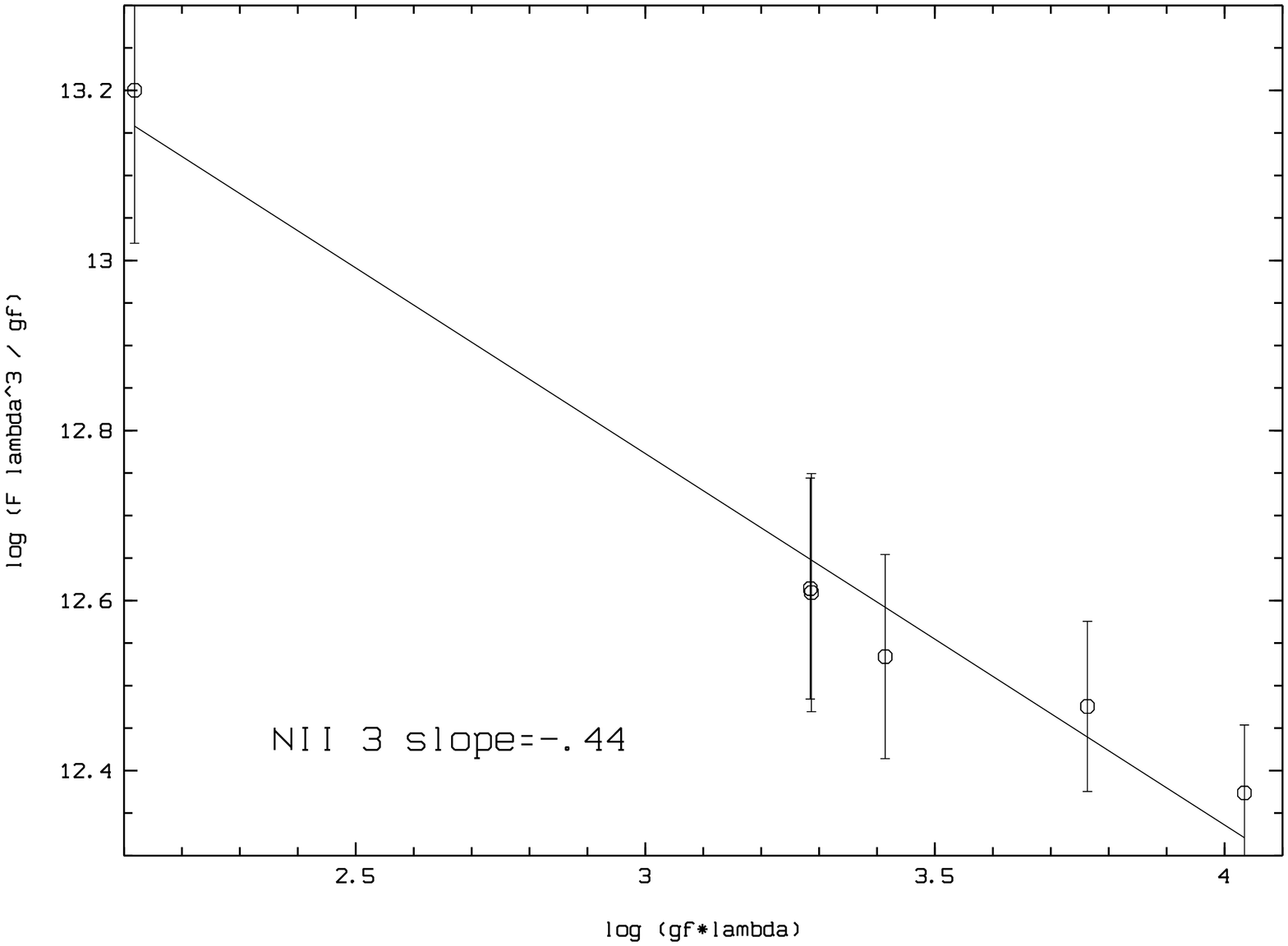}
\includegraphics[angle=-90,width=8.5cm,bb= 55 100 530 780,clip]{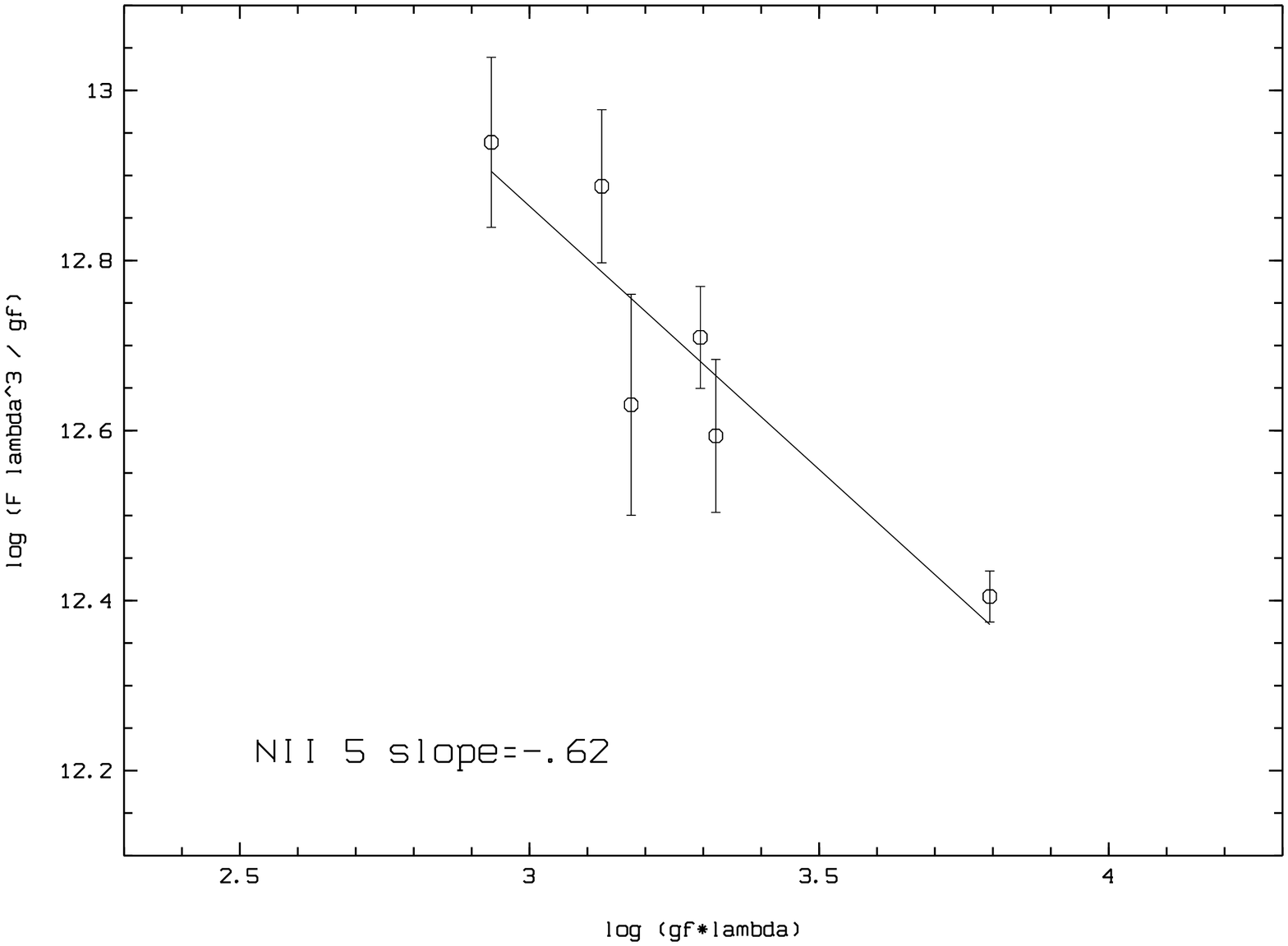}
\includegraphics[angle=-90,width=8.5cm,bb= 55 100 530 780,clip]{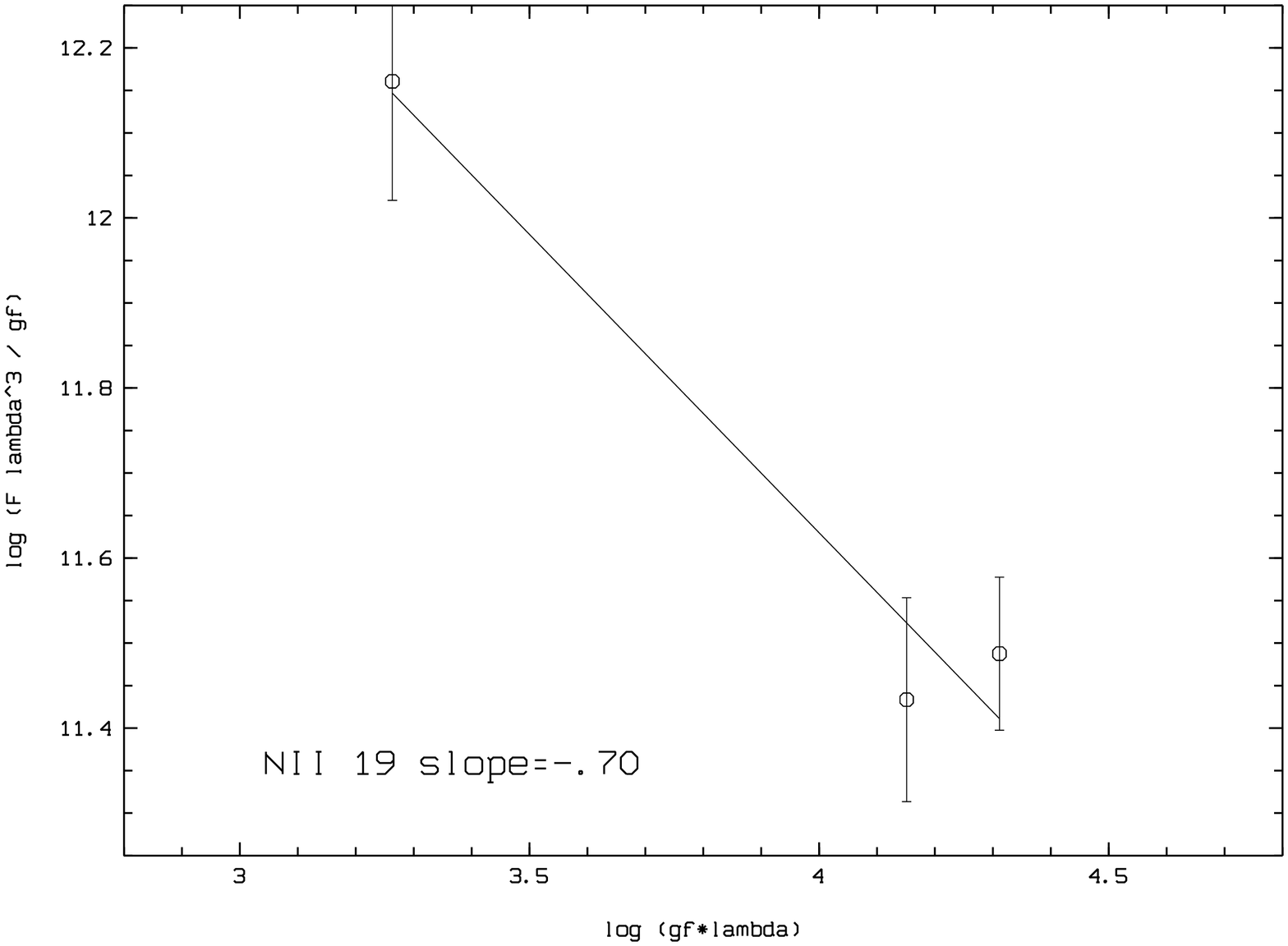}
\includegraphics[angle=-90,width=8.5cm,bb= 55 100 530 780,clip]{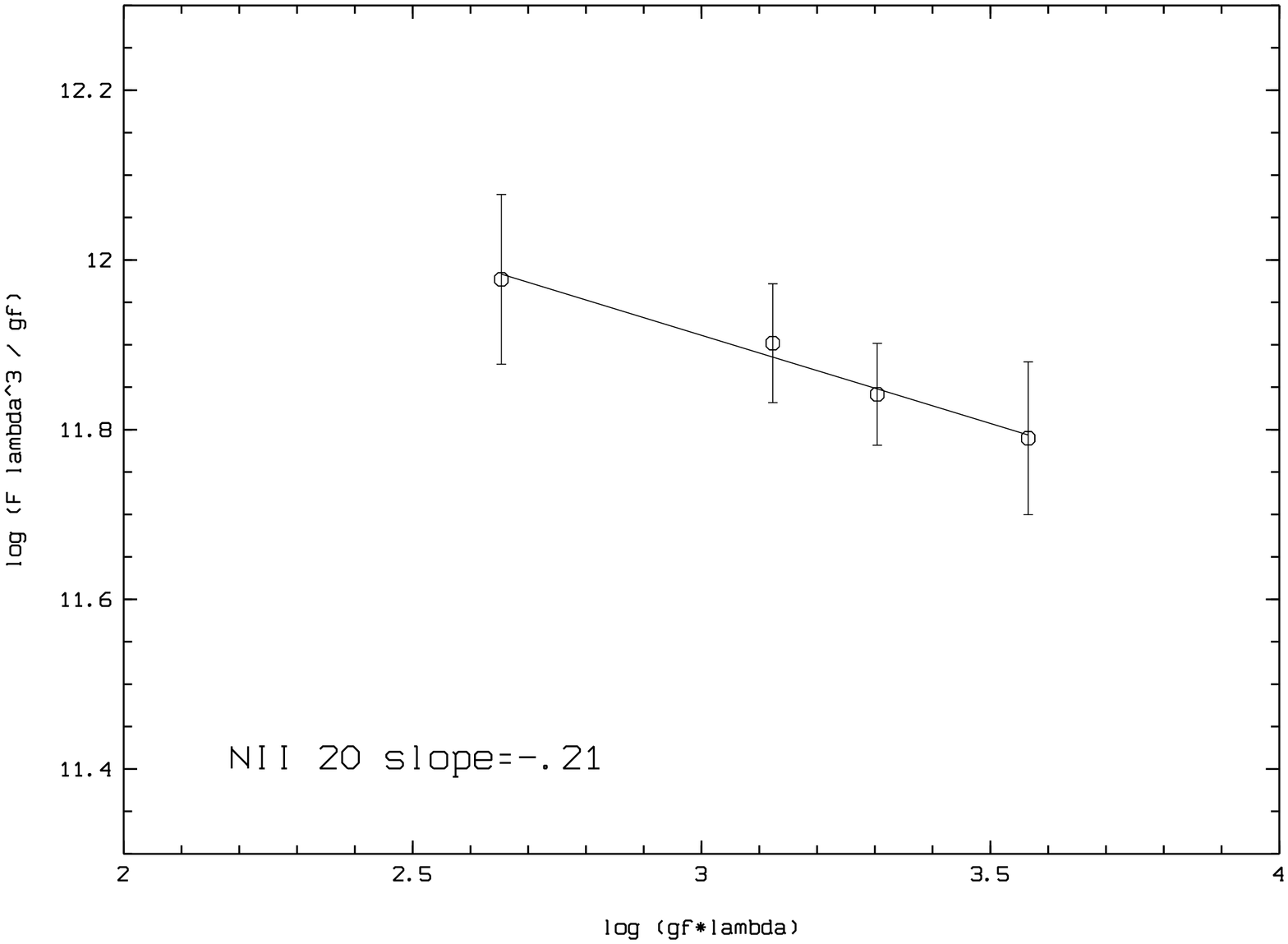}
\end{figure}

\begin{figure}
\centering
\includegraphics[angle=-90,width=8.5cm,bb= 55 100 530 780,clip]{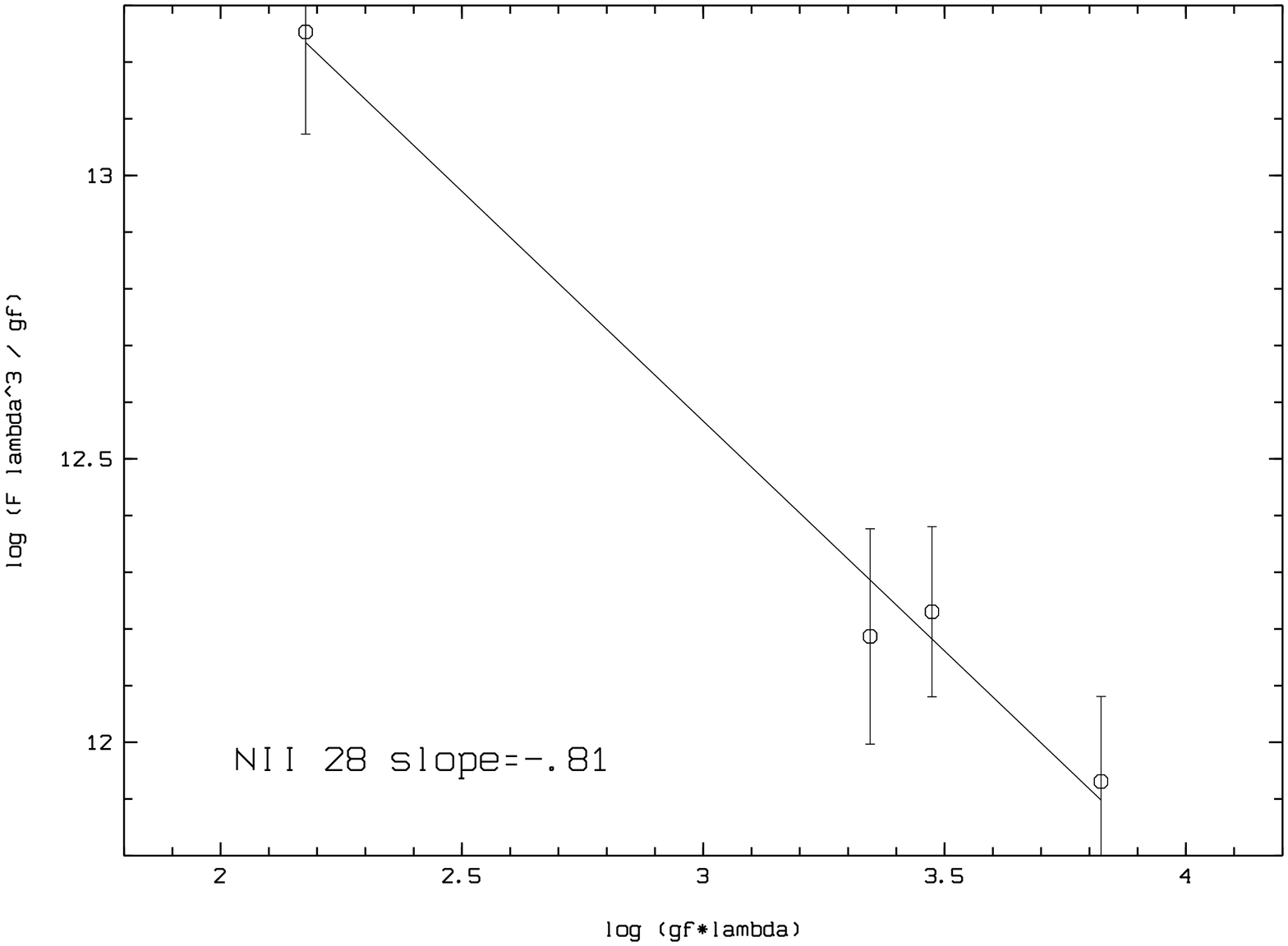}
\includegraphics[angle=-90,width=8.5cm,bb= 55 100 530 780,clip]{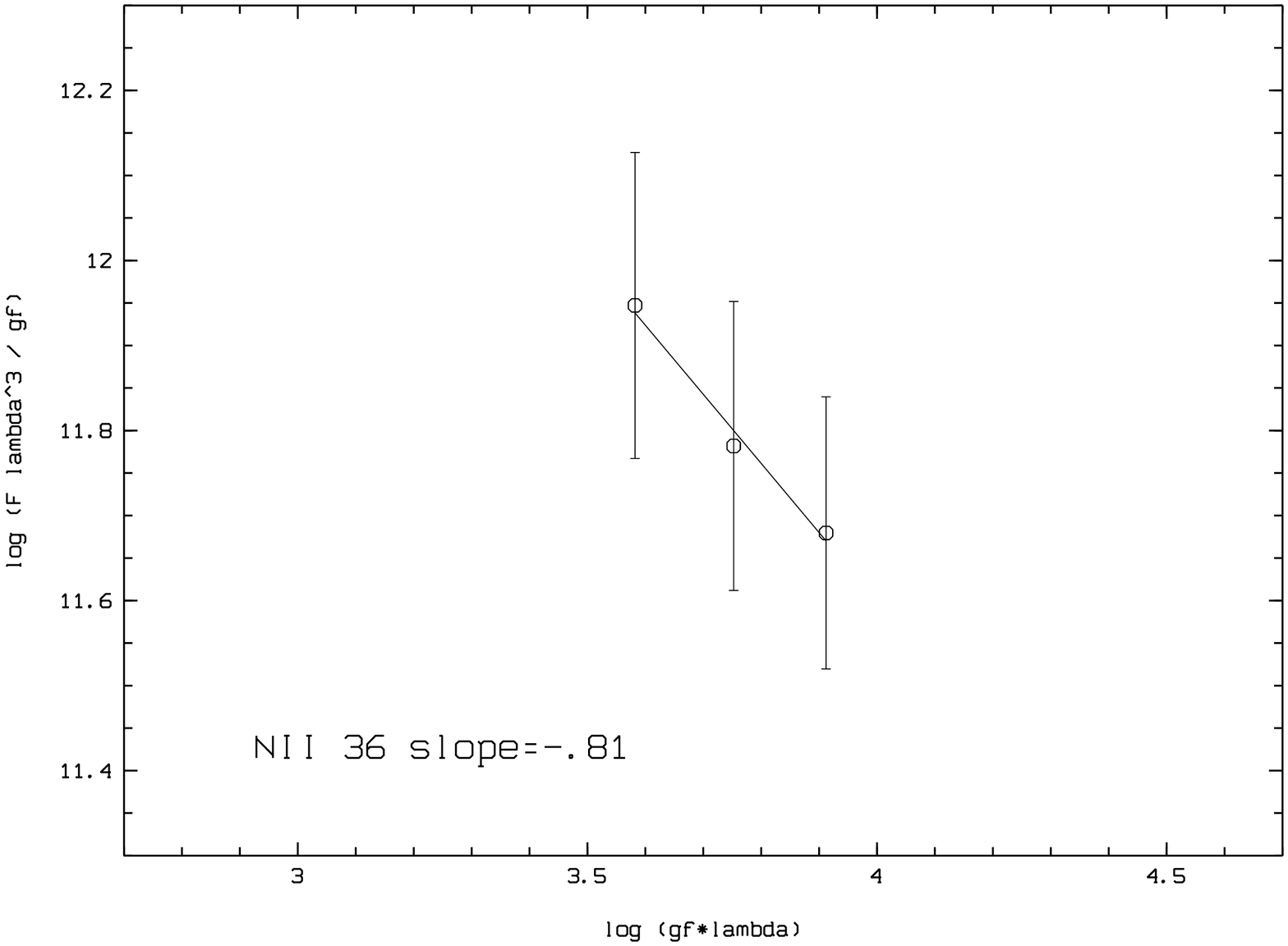}
\includegraphics[angle=-90,width=8.5cm,bb= 55 100 530 780,clip]{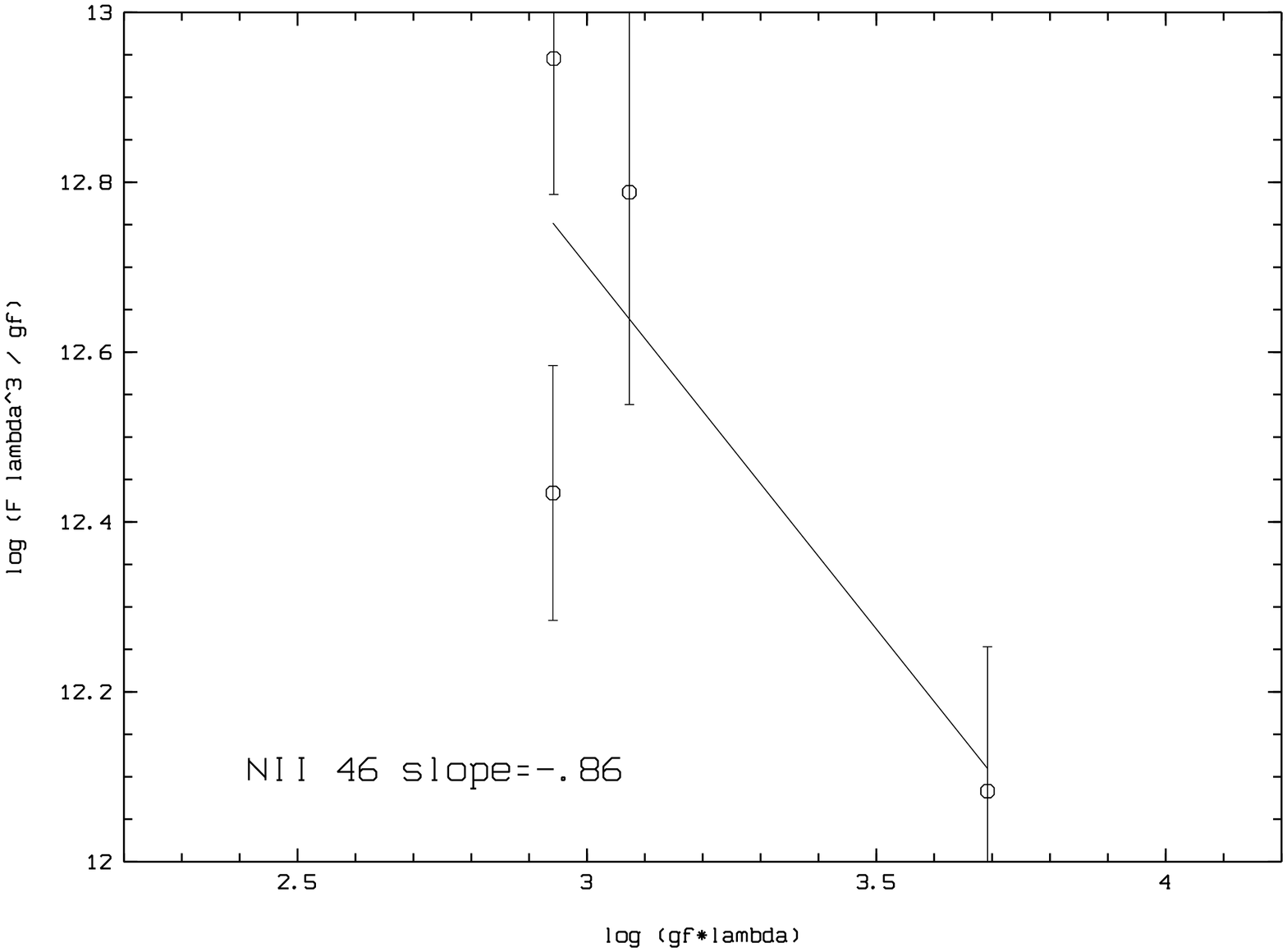}
\caption{Plots of $\log\frac{{F\lambda^{3}}}{{gf}}$ against $\log(gf\lambda)$  (SAC) for NII multiplets 3,5,19,20,28,36,46}
\label{Fig3}
\end{figure}

\section{Discussion}

The origin of the emission spectrum of \object{P Cygni} was partly discussed
by Wolf and Stahl (\cite{W85}) who suggested that the upper terms of
multiplets 115 and 117 of Fe III were pumped by two lines of He I. Pumping by
two UV transitions in He I ($\lambda\lambda$ 522, 537), followed by downward
cascades, was suggested by Markova and de Groot (\cite{MG97}) to explain the
Fe III emission lines of multiplets 113, 114, 118 and 119. In addition, the
authors noted that the N II multiplets with the highest upper term excitation
potential ($\sim$25 eV) might be populated by dielectronic recombination,
but the temperature of the wind seems to be  too low for this mechanism to
work. Stahl et al (\cite{S93}) however suppose that the highest 
excitation N II multiplets are excited by an unknown pumping mechanism.

The optical thickness of the more excited Fe III and \\ N II lines without
blueshifted absorption components, established by us, is quite surprising,
while the indications of an increase of SAC slope with increasing excitation,
if confirmed in future work, would be in fact counter-intuitive. At the same
log$(gf \lambda)$ more excited lines are expected to be in any case optically
thinner. A conspirancy of differential non-LTE effects appears
unlikely as an explanation, especially when the points for a multiplet do not
show large deviations from a curve. However we should note, that 
it is hard to rigorously test for such an effect, as both the upper and  
lower levels of the observed optical Fe III and N II lines could in principle
show deviations from LTE. In objects showing many Fe II emission lines, the
lower levels of strong lines are very often in LTE and in that case one only
needs to test for effects depending on the upper level.

Another explanation, which could be suggested, will not work either, if the
increase of curve slope at the same log$(gf \lambda)$ for more excited
multiplets is confirmed in future work. One might suppose that the more
excited lines are only emitted from certain parts of a spherically symmetric
wind. In that case the absence of blueshifted absorption components for
excited multiplets might be explained by a large source function, so the
emission in the lines per unit surface area, would be larger than that 
of the photosphere. In that case extra emission instead of absorption would be
present on the blue side of the more excited lines, resulting in a 
blueshift of the mean radial velocity, as was in fact observed by Markova
and de Groot (\cite{MG97}). Such a blueshift could be particularly important
if part of the receding material was behind and so occulted by the  
photosphere, but the symmetry of the lines, noted by Markova and de Groot,
would suggest that such occultation is small. In any case if the wind has
spherical symmetry, the less excited lines could not be optically thinner,
because those lines would also be emitted in the region of emission  
of the more excited lines, where they should not be optically thinner than
those excited lines.

If we exclude the previous explanations, we may need to conclude that the 
wind of P Cyg is anisotropic. Optically thick excited lines with no blue 
shifted absorption components would in that case be mainly formed in clouds,
which either have a large source function or which do not occult much 
of the photosphere. The optically thinner lower excitation lines with
blueshifted absorption, could then be mainly formed in large regions of the 
wind, covering the photosphere, with only a small optically thicker
contribution to their emission from the clouds. The physical reason for such
a situation is however not immediately clear. Heating due to shock waves might
occur in the clouds, producing extra ionizing radiation, followed by  
more recombinations and cascades plus pumping to extremely excited levels 
Such a situation might also be a way of explaining the presence of the
most excited N II lines. Theoretical work is required to see whether this is a
viable option.

We must also finally point out that effects such as an increase in
self-absorption curve slope for more excited multiplets, would be hard to 
find in detailed spectral syntheses, based on spherically symmetric wind
models. Semi-empirical methods can be quite powerful in searches for unknown
physical processes, present in relatively complex situations.

\begin{acknowledgements}
N. Markova is grateful for the warm hospitality of the Observatoire de
Marseille (France). G. Muratorio and M. Friedjung wish to acknowledge the
hospitality of the National Astronomical Observatory (NAO, Bulgaria), where
part of the work was done. This study was supported by the French CNRS and the
Bulgarian Academy of Sciences. Research at the NAO was in part supported by
the NSF through grant F-813/1998 to the Bulgarian Ministry of Education.
\end{acknowledgements}
{}

\end{document}